\documentclass[aps,prd,twocolumn,showpacs,superscriptaddress,groupedaddress]{revtex4-1}  % for review and submission
\RequirePackage{xspace}
\usepackage{graphicx}  % needed for figures
\usepackage{dcolumn}   % needed for some tables
\usepackage{bm}        % for math
\usepackage{amssymb}   % for math
\usepackage{amsmath}
\usepackage{lineno}
\usepackage{multirow}
\usepackage{color}
\usepackage{hyperref}
\usepackage{url}
\begin{document}

%Symbols
%%%%%%%%%%%%%%%%%   LEPTONS   %%%%%%%%%%%%%%%%%
%%%%%%%%%%%%%%%%%%%%%%%%%%%%%%%%%%%%%%%%%%%%%%%
\let\emi\en

\def\nue        {\ensuremath{\nu_e}\xspace}
\def\num        {\ensuremath{\nu_{\mu}}\xspace}
\def\nut        {\ensuremath{\nu_\tau}\xspace}

%%%%%%%%%%%%%%%%%   SME Specific   %%%%%%%%%%%%%%%%%
%%%%%%%%%%%%%%%%%%%%%%%%%%%%%%%%%%%%%%%%%%%%%%%

%\def\somega     {\ensuremath{\omega_{\oplus}}\xspace}
%\def\sday       {\ensuremath{23^{h}56^{m}4.0916^{s}}\xspace}
%%\def\Cab        {\ensuremath{\mathcal{C}_{ab}}\xspace}
%%\def\As         {\ensuremath{(\mathcal{A}_{s})_{ab}}\xspace}
%%\def\Ac         {\ensuremath{(\mathcal{A}_{c})_{ab}}\xspace}
%%\def\Bs         {\ensuremath{(\mathcal{B}_{s})_{ab}}\xspace}
%%\def\Bc         {\ensuremath{(\mathcal{B}_{c})_{ab}}\xspace}
%%\def\al         {\ensuremath{(a_{L})_{ab}}\xspace}
%%\def\cl         {\ensuremath{(c_{L})_{ab}}\xspace}

\def\somega     {\ensuremath{\omega_{\oplus}}}
\def\sday       {\ensuremath{23^{h}56^{m}4.0916^{s}}}
\def\Ce        {\ensuremath{\mathcal{C}_{\mu e}}}
\def\Ase         {\ensuremath{(\mathcal{A}_{s})_{\mu e}}}
\def\Ace         {\ensuremath{(\mathcal{A}_{c})_{\mu e}}}
\def\Bse         {\ensuremath{(\mathcal{B}_{s})_{\mu e}}}
\def\Bce         {\ensuremath{(\mathcal{B}_{c})_{\mu e}}}
\def\Ctau        {\ensuremath{\mathcal{C}_{\mu \tau}}}
\def\Astau         {\ensuremath{(\mathcal{A}_{s})_{\mu \tau}}}
\def\Actau         {\ensuremath{(\mathcal{A}_{c})_{\mu \tau}}}
\def\Bstau         {\ensuremath{(\mathcal{B}_{s})_{\mu \tau}}}
\def\Bctau         {\ensuremath{(\mathcal{B}_{c})_{\mu \tau}}}

\def\Cab        {\ensuremath{\mathcal{C}_{ab}}}
\def\Asab         {\ensuremath{(\mathcal{A}_{s})_{ab}}}
\def\Acab         {\ensuremath{(\mathcal{A}_{c})_{ab}}}
\def\Bsab         {\ensuremath{(\mathcal{B}_{s})_{ab}}}
\def\Bcab         {\ensuremath{(\mathcal{B}_{c})_{ab}}}
\def\al         {\ensuremath{(a_{L})_{ab}}}
\def\cl         {\ensuremath{(c_{L})_{ab}}}

\def\Cmb        {\ensuremath{\mathcal{C}_{\mu b}}}
\def\Asmb         {\ensuremath{(\mathcal{A}_{s})_{\mu b}}}
\def\Acmb         {\ensuremath{(\mathcal{A}_{c})_{\mu b}}}
\def\Bsmb         {\ensuremath{(\mathcal{B}_{s})_{\mu b}}}
\def\Bcmb         {\ensuremath{(\mathcal{B}_{c})_{\mu b}}}
\def\alm         {\ensuremath{(a_{L})_{\mu b}}}
\def\clm         {\ensuremath{(c_{L})_{\mu b}}}

% The following information is for internal review, please remove them for submission
%\widetext
%\leftline{Version 1 as of \today}
%\leftline{Primary authors: T2K}
%\leftline{To be submitted to PRD}
%\leftline{Comment to {\tt d0-run2eb-nnn@fnal.gov} by xxx, yyy}
%\centerline{\em D\O\ INTERNAL DOCUMENT -- NOT FOR PUBLIC DISTRIBUTION}

\title{Search for Lorentz and CPT violation using sidereal time dependence of neutrino flavor transitions over a short baseline}

%%%%%%%%%%%%%%%%%%%%%%%%%%%%%%%%%%%%%%%%%%%%%%%%%%%%%%%%%%%%%%
% T2K author list generated on Thu, 16 Feb 2017 01:35:48 +0900
% setting: extra = 0 revtex = 1 ptep = 0 simple = 0 xml = 0 yearrule = 1 shiftrule = 1
%         author list from archive (starting 2016/03/29 until now)
% Number of authors = 357
%%%%%%%%%%%%%%%%%%%%%%%%%%%%%%%%%%%%%%%%%%%%%%%%%%%%%%%%%%%%%%

\newcommand{\INSTEE}{\affiliation{University of Bern, Albert Einstein Center for Fundamental Physics, Laboratory for High Energy Physics (LHEP), Bern, Switzerland}}
\newcommand{\INSTFE}{\affiliation{Boston University, Department of Physics, Boston, Massachusetts, U.S.A.}}
\newcommand{\INSTD}{\affiliation{University of British Columbia, Department of Physics and Astronomy, Vancouver, British Columbia, Canada}}
\newcommand{\INSTGA}{\affiliation{University of California, Irvine, Department of Physics and Astronomy, Irvine, California, U.S.A.}}
\newcommand{\INSTI}{\affiliation{IRFU, CEA Saclay, Gif-sur-Yvette, France}}
\newcommand{\INSTGB}{\affiliation{University of Colorado at Boulder, Department of Physics, Boulder, Colorado, U.S.A.}}
\newcommand{\INSTFG}{\affiliation{Colorado State University, Department of Physics, Fort Collins, Colorado, U.S.A.}}
\newcommand{\INSTFH}{\affiliation{Duke University, Department of Physics, Durham, North Carolina, U.S.A.}}
\newcommand{\INSTBA}{\affiliation{Ecole Polytechnique, IN2P3-CNRS, Laboratoire Leprince-Ringuet, Palaiseau, France }}
\newcommand{\INSTEF}{\affiliation{ETH Zurich, Institute for Particle Physics, Zurich, Switzerland}}
\newcommand{\INSTEG}{\affiliation{University of Geneva, Section de Physique, DPNC, Geneva, Switzerland}}
\newcommand{\INSTDG}{\affiliation{H. Niewodniczanski Institute of Nuclear Physics PAN, Cracow, Poland}}
\newcommand{\INSTCB}{\affiliation{High Energy Accelerator Research Organization (KEK), Tsukuba, Ibaraki, Japan}}
\newcommand{\INSTED}{\affiliation{Institut de Fisica d'Altes Energies (IFAE), The Barcelona Institute of Science and Technology, Campus UAB, Bellaterra (Barcelona) Spain}}
\newcommand{\INSTEC}{\affiliation{IFIC (CSIC \& University of Valencia), Valencia, Spain}}
\newcommand{\INSTEI}{\affiliation{Imperial College London, Department of Physics, London, United Kingdom}}
\newcommand{\INSTGF}{\affiliation{INFN Sezione di Bari and Universit\`a e Politecnico di Bari, Dipartimento Interuniversitario di Fisica, Bari, Italy}}
\newcommand{\INSTBE}{\affiliation{INFN Sezione di Napoli and Universit\`a di Napoli, Dipartimento di Fisica, Napoli, Italy}}
\newcommand{\INSTBF}{\affiliation{INFN Sezione di Padova and Universit\`a di Padova, Dipartimento di Fisica, Padova, Italy}}
\newcommand{\INSTBD}{\affiliation{INFN Sezione di Roma and Universit\`a di Roma ``La Sapienza'', Roma, Italy}}
\newcommand{\INSTEB}{\affiliation{Institute for Nuclear Research of the Russian Academy of Sciences, Moscow, Russia}}
\newcommand{\INSTHA}{\affiliation{Kavli Institute for the Physics and Mathematics of the Universe (WPI), The University of Tokyo Institutes for Advanced Study, University of Tokyo, Kashiwa, Chiba, Japan}}
\newcommand{\INSTCC}{\affiliation{Kobe University, Kobe, Japan}}
\newcommand{\INSTCD}{\affiliation{Kyoto University, Department of Physics, Kyoto, Japan}}
\newcommand{\INSTEJ}{\affiliation{Lancaster University, Physics Department, Lancaster, United Kingdom}}
\newcommand{\INSTFC}{\affiliation{University of Liverpool, Department of Physics, Liverpool, United Kingdom}}
\newcommand{\INSTFI}{\affiliation{Louisiana State University, Department of Physics and Astronomy, Baton Rouge, Louisiana, U.S.A.}}
\newcommand{\INSTJ}{\affiliation{Universit\'e de Lyon, Universit\'e Claude Bernard Lyon 1, IPN Lyon (IN2P3), Villeurbanne, France}}
\newcommand{\INSTHB}{\affiliation{Michigan State University, Department of Physics and Astronomy,  East Lansing, Michigan, U.S.A.}}
\newcommand{\INSTCE}{\affiliation{Miyagi University of Education, Department of Physics, Sendai, Japan}}
\newcommand{\INSTDF}{\affiliation{National Centre for Nuclear Research, Warsaw, Poland}}
\newcommand{\INSTFJ}{\affiliation{State University of New York at Stony Brook, Department of Physics and Astronomy, Stony Brook, New York, U.S.A.}}
\newcommand{\INSTGJ}{\affiliation{Okayama University, Department of Physics, Okayama, Japan}}
\newcommand{\INSTCF}{\affiliation{Osaka City University, Department of Physics, Osaka, Japan}}
\newcommand{\INSTGG}{\affiliation{Oxford University, Department of Physics, Oxford, United Kingdom}}
\newcommand{\INSTBB}{\affiliation{UPMC, Universit\'e Paris Diderot, CNRS/IN2P3, Laboratoire de Physique Nucl\'eaire et de Hautes Energies (LPNHE), Paris, France}}
\newcommand{\INSTGC}{\affiliation{University of Pittsburgh, Department of Physics and Astronomy, Pittsburgh, Pennsylvania, U.S.A.}}
\newcommand{\INSTFA}{\affiliation{Queen Mary University of London, School of Physics and Astronomy, London, United Kingdom}}
\newcommand{\INSTE}{\affiliation{University of Regina, Department of Physics, Regina, Saskatchewan, Canada}}
\newcommand{\INSTGD}{\affiliation{University of Rochester, Department of Physics and Astronomy, Rochester, New York, U.S.A.}}
\newcommand{\INSTHC}{\affiliation{Royal Holloway University of London, Department of Physics, Egham, Surrey, United Kingdom}}
\newcommand{\INSTBC}{\affiliation{RWTH Aachen University, III. Physikalisches Institut, Aachen, Germany}}
\newcommand{\INSTFB}{\affiliation{University of Sheffield, Department of Physics and Astronomy, Sheffield, United Kingdom}}
\newcommand{\INSTDI}{\affiliation{University of Silesia, Institute of Physics, Katowice, Poland}}
\newcommand{\INSTEH}{\affiliation{STFC, Rutherford Appleton Laboratory, Harwell Oxford,  and  Daresbury Laboratory, Warrington, United Kingdom}}
\newcommand{\INSTCH}{\affiliation{University of Tokyo, Department of Physics, Tokyo, Japan}}
\newcommand{\INSTBJ}{\affiliation{University of Tokyo, Institute for Cosmic Ray Research, Kamioka Observatory, Kamioka, Japan}}
\newcommand{\INSTCG}{\affiliation{University of Tokyo, Institute for Cosmic Ray Research, Research Center for Cosmic Neutrinos, Kashiwa, Japan}}
\newcommand{\INSTGI}{\affiliation{Tokyo Metropolitan University, Department of Physics, Tokyo, Japan}}
\newcommand{\INSTF}{\affiliation{University of Toronto, Department of Physics, Toronto, Ontario, Canada}}
\newcommand{\INSTB}{\affiliation{TRIUMF, Vancouver, British Columbia, Canada}}
\newcommand{\INSTG}{\affiliation{University of Victoria, Department of Physics and Astronomy, Victoria, British Columbia, Canada}}
\newcommand{\INSTDJ}{\affiliation{University of Warsaw, Faculty of Physics, Warsaw, Poland}}
\newcommand{\INSTDH}{\affiliation{Warsaw University of Technology, Institute of Radioelectronics, Warsaw, Poland}}
\newcommand{\INSTFD}{\affiliation{University of Warwick, Department of Physics, Coventry, United Kingdom}}
\newcommand{\INSTGE}{\affiliation{University of Washington, Department of Physics, Seattle, Washington, U.S.A.}}
\newcommand{\INSTGH}{\affiliation{University of Winnipeg, Department of Physics, Winnipeg, Manitoba, Canada}}
\newcommand{\INSTEA}{\affiliation{Wroclaw University, Faculty of Physics and Astronomy, Wroclaw, Poland}}
\newcommand{\INSTHE}{\affiliation{Yokohama National University, Faculty of Engineering, Yokohama, Japan}}
\newcommand{\INSTH}{\affiliation{York University, Department of Physics and Astronomy, Toronto, Ontario, Canada}}

\INSTEE
\INSTFE
\INSTD
\INSTGA
\INSTI
\INSTGB
\INSTFG
\INSTFH
\INSTBA
\INSTEF
\INSTEG
\INSTDG
\INSTCB
\INSTED
\INSTEC
\INSTEI
\INSTGF
\INSTBE
\INSTBF
\INSTBD
\INSTEB
\INSTHA
\INSTCC
\INSTCD
\INSTEJ
\INSTFC
\INSTFI
\INSTJ
\INSTHB
\INSTCE
\INSTDF
\INSTFJ
\INSTGJ
\INSTCF
\INSTGG
\INSTBB
\INSTGC
\INSTFA
\INSTE
\INSTGD
\INSTHC
\INSTBC
\INSTFB
\INSTDI
\INSTEH
\INSTCH
\INSTBJ
\INSTCG
\INSTGI
\INSTF
\INSTB
\INSTG
\INSTDJ
\INSTDH
\INSTFD
\INSTGE
\INSTGH
\INSTEA
\INSTHE
\INSTH

\author{K.\,Abe}\INSTBJ
\author{J.\,Amey}\INSTEI
\author{C.\,Andreopoulos}\INSTEH\INSTFC
\author{M.\,Antonova}\INSTEB
\author{S.\,Aoki}\INSTCC
\author{A.\,Ariga}\INSTEE
\author{S.\,Assylbekov}\INSTFG
\author{D.\,Autiero}\INSTJ
\author{S.\,Ban}\INSTCD
\author{F.C.T.\,Barbato}\INSTBE
% ...

\collaboration{The T2K Collaboration}\noaffiliation

\begin{abstract}
A class of extensions of the Standard Model allows Lorentz and CPT violations, which can be identified by the observation of sidereal modulations in the neutrino interaction rate.
A search for such modulations was performed using the T2K on-axis near detector. Two complementary methods were used in this study, both of which resulted in no evidence of a signal. Limits on associated Lorentz and CPT violating terms from the Standard Model Extension have been derived taking into account their correlations in this model for the first time. These results imply such symmetry violations are suppressed by a factor of more than $10^{20}$ at the GeV scale.
\end{abstract}
\maketitle
\clearpage
\newpage

\section{Introduction}

While Lorentz invariance is a cornerstone of the Standard Model (SM) of particle physics, violations of this symmetry 
are allowed in a variety of models~\cite{kostelecky1989spontaneous,hawking1976breakdown,hinchliffe2004review}
at or around the Planck scale, $m_{P} \sim 10^{19}~$GeV. 
At energies relevant to modern experiments, Lorentz invariance violating (LV) processes are 
expected to be suppressed at least by $\sim 1/m_{P}$.
Experimental observations of such phenomena would provide direct access to physics at the 
Planck scale and precision tests have been 
performed to overcome this suppression (c.f.~\cite{kostelecky2011data} for a review).
Neutrino oscillations can be used as a natural interferometer to probe even weak departures from 
this symmetry and have been studied with 
accelerator~\cite{Auerbach:2005tq, AguilarArevalo:2011yi, Adamson:2008aa, Adamson:2012hp,Adamson:2010rn,rebel2013search},
reactor~\cite{Abe:2012gw}, 
and atmospheric~\cite{Abbasi:2010kx,Abe:2014wla} neutrinos. 

Lorentz and charge-parity-time (CPT) symmetry violations can be described within the 
context of the standard model extension (SME)~\cite{colladay1998lorentz}, an observer-independent effective field 
theory that incorporates all possible spontaneous LV operators with the SM Lagrangian.
In general the SME allows two classes of effects for neutrino oscillations, 
sidereal violations, in which the presence of a preferred spatial direction induces 
oscillation effects that vary with the neutrino travel direction, and 
spectral anomalies~\cite{Kostelecky:2011gq,Kostelecky:2003xn,Kostelecky:2003cr}.
For a terrestrial fixed-baseline experiment, the rotation of the Earth induces a change in the 
direction of the neutrino target-detector vector relative to a fixed coordinate system such 
that a LV signal of the former type would manifest itself as a variation in the neutrino oscillation probability with 
sidereal time.

This paper reports on a search for evidence of sidereal-dependent $\nu_{\mu}$ disappearance over an average baseline of 233.6~m using the T2K experiment.  
After introducing Lorentz invariance violating oscillations within the SME and describing the T2K experiment, 
the selection of an analysis sample composed predominately of muon neutrinos inside the INGRID~\cite{abe2011t2k,abe2012measurements} detector is presented.
Results of two complementary analyses of the data and concluding remarks follow thereafter.
 
\section{LV effects on neutrino oscillations at short distances}
In this analysis, the LV is probed through $\num$ disappearance channel. In the SME framework, the disappearance probability of a $\num$ over short baselines is given by~\cite{Kostelecky:2003xn}:
\begin{eqnarray}
\begin{aligned}
      P_{\num \rightarrow \num} = & 1 - \sum_{b, b \neq \mu} \frac{L^{2}}{(\hbar c)^{2}} \bigl| \Cmb  \\ 
        &+ \Asmb \sin(\somega T_{\oplus}) + \Acmb\cos(\somega T_{\oplus}) \\
        &+ \Bsmb   \sin(2\somega T_{\oplus}) + \Bcmb\cos(2\somega T_{\oplus})\bigr|^{2},
%%\nonumber
\label{eq:shortprob}
\end{aligned}
\end{eqnarray}
\noindent where $L$ is the distance travelled before detection. Equation~(\ref{eq:shortprob}) is valid as long as $L \ll L_{osc}$, where $L_{osc}$ is the typical distance of standard $\num \rightarrow \nu_{b}$ oscillations~\cite{Agashe:2014kda}.
$T_{\oplus}$ is the local sidereal time and $ \somega = \frac{2\pi}{\sday}$ is the Earth's sidereal frequency. Under a three flavour neutrino hypothesis, oscillations of $\num$ to $\nue$ and $\nut$ can occur.
In general, the ten coefficients $\Cmb$, $\Acmb$, $\Asmb$, $\Bcmb$, and $\Bsmb$ ($b=e,\tau$) are functions of the neutrino energy $E$, the neutrino beam direction at the time origin (see below), and 
of forty parameters within the SME which carry explicit Lorentz and CPT violation information:
$(a_{L})^{\alpha}_{\mu b}$ and $(c_{L})^{\alpha \beta}_{\mu b}$ ($b=e,\tau$)~\cite{diaz2009perturbative}. The $(a_{L})^{\alpha}_{\mu b}$ ($(c_{L})^{\alpha \beta}_{\mu b}$) are constant coefficients associated with CPT odd (even) vector (tensor) fields.
It should be noted that the impact of $(a_{L})^{\alpha}_{ab}$ and $(c_{L})^{\alpha \beta}_{ab}$ on the set of ten coefficients depend on the absolute direction of the neutrino baseline~\cite{diaz2009perturbative}.
In the analysis to follow, a search for a sidereal variations is performed relative to an inertial frame 
centered on the Sun assuming it to be stationary during the data taking period.
Other than the choice of the origin of the time coordinate, this frame is the same as in~\cite{Katori:2012pe}. The time origin $T=0$ is chosen as 1 January 1970, 09:00:00 Coordinated Universal Time. Data will be studied using the local sidereal phase (LSP), which is defined as $\text{LSP}=\mbox{mod}( T_{\oplus} \somega / 2\pi )$.

\section{Experimental setup}

The T2K long-baseline neutrino experiment uses the collision of 30~GeV protons from the Japan Proton Accelerator Research Complex (J-PARC) with a graphite target, and focuses charged mesons produced in the subsequent interactions 
along the primary proton beam direction using a series of magnetic horns.
Downstream of the production target is a 96~m long decay volume in which these mesons 
decay to produce a beam of primarily muon neutrinos ($99.3\%$ $\nu_{\mu} + \overline{\nu}_{\mu}$ along the beam axis).  

This study is based on data accumulated from 2010 to 2013, divided into four run periods, and corresponds to $6.63\times 10^{20}$ protons on target (POT) exposure of the INGRID detector in neutrino-mode.
%The neutrino beamline is located at the coast of the Pacific ocean.
The neutrino beam is defined by the beam colatitude $\chi=53.55087^{\circ}$ in the Earth-centered frame with the same fixed axis than the Sun-centered frame. At the beamline location, a local frame is defined where the z-axis corresponds to the zenith. The beam direction in this local frame is defined by the zenith angle $\theta=93.637^{\circ}$ and at the azimuthal angle $\phi=270.319^{\circ}$. A more detailed description of the T2K experiment can be found in~\cite{abe2011t2k}.

The INGRID detector is located $280$~m downstream of the graphite target and is composed of 
14 $120$~cm$ \times 120$~cm$ \times 109~$cm modules assembled in a cross-shaped structure. 
Each module holds 11 tracking segments built from pairs of orthogonally oriented scintillator planes 
interleaved with nine iron planes.
The scintillator planes are built from 24 plastic scintillator bars connected to 
multi-pixel photon counters (MPPCs). Situated on the beam center, INGRID high event rate makes it well suited to a search for a sidereal variation in the $\nu_{\mu}$ interactions.

\begin{figure}[ht]
  \centering
  \includegraphics[width=.48\textwidth]{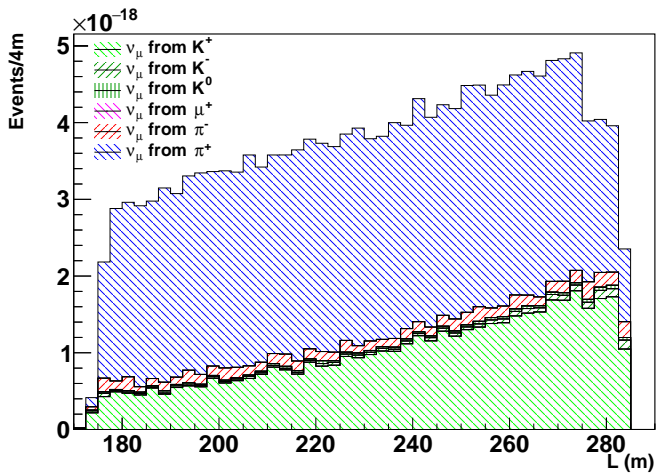}
  \caption{\label{fig:nuZ} Flight length to the INGRID detector for MC $\nu_{\mu}$ produced in the T2K decay volume. 
                           The distribution is separated based on the neutrino's parent particle. }
\end{figure}

Although the $\nu_{\mu}\rightarrow \nu_{\mu}$ oscillation probability in Equation~(\ref{eq:shortprob}) depends on the 
square of the neutrino flight length, the precise distance from creation to detection for each neutrino is unknown.
Indeed, the neutrino's parent meson may decay anywhere along the decay volume as shown in Figure~\ref{fig:nuZ}. 
As a result the present analysis uses the mean of this distribution, $L_{ave} = 233.6$~m, 
as an effective distance travelled for all candidate events. 
Similarly, the mean neutrino energy of the flux at the INGRID detector, $E_{ave} = 2.7~\mbox{GeV}$, is used.
%%%%%
%%
%  Neutrino Flight length to INGRID
%%
%%%%
%%%%%%%%%%%%%

\section{$\nu_{\mu}$ Event Selection and Systematic Uncertainties}
\label{sec:Selection}

\subsection{The INGRID $\nu_{\mu}$ event selection}
\label{sec:Selection}

\begin{figure}[hbt]
  \centering
  \includegraphics[width=.48\textwidth]{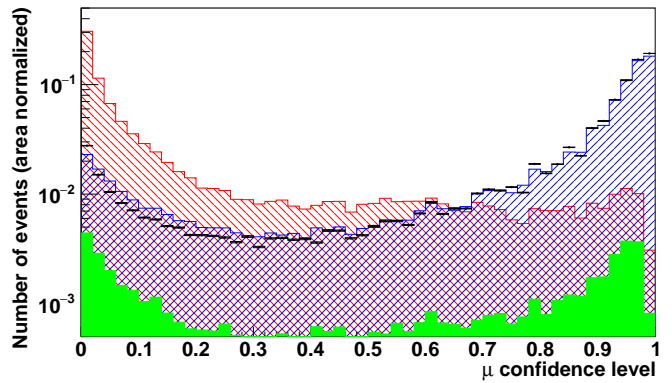}
  \caption{ Distribution of the $\mu_{CL}$ variable for $\nu_{\mu}$ CC (blue), $\nu_{e}$ CC (red), and NC events (green) from the MC are overlaid with data (black). The data, $\nu_{\mu}$ CC and $\nu_{\mu}$ NC histograms are first normalized by protons on target, and then, scaled by one over the number of $\nu_{\mu}$ CC events to preserve their relative proportions. The $\nu_{e}$ CC histogram is area normalized to compare with the $\nu_{\mu}$ CC histogram. The pink arrow represents the lower cut value on the $\mu_{CL}$ that defines the $\num$ event selection.}
\label{Likelihood} 
\end{figure}

To prevent LV oscillation-induced $\nu_{e}$ and $\nu_{\tau}$ from washing out an LV effect on the $\nu_{\mu}$ data, it is essential to select a sample with very high $\nu_{\mu}$ purity. Since the $\nu_{\tau}$ CC interactions have a 3.5~GeV production threshold, their cross section in the T2K energy range is very small. Their impact on the analysis was evaluated to be negligible. Consequently, no attempts were made to further reject them in the signal selection.\\ 
%%
%Even the $\sim 0.7 \%$ contamination of $\nu_{e}$ intrinsic to the beam itself may suppress a weak signal so an optimized event selection has been developed to reduce these backgrounds.
%%
Charged-current neutrino $\nu_{\mu}$ interactions within INGRID are identified by a reconstructed track 
consistent with a muon originating in the detector fiducial volume, and coincident in time with the expected arrival of neutrinos in the beam originated from a given proton bunch.
In addition to a set of cuts to define a basic lepton-like sample~\cite{abe2014measurement}, a 
likelihood function, hereafter referred to as muon confidence level ($\mu_{CL}$), is used to further 
separate tracks produced by muons from showers produced by electrons or hadrons. 
This function is based on four discriminating variables: the number of active scintillator bars 
transverse to the beam direction averaged over the number of active planes, \textit{i.e.} planes having at least one hit belonging to the track; 
the primary track's length;
the dispersion of the track's energy deposition with distance; 
and the number of active scintillator bars close to the primary interaction vertex. 
The first three variables focus on the tendency for showers to have a broader transverse development 
and varying rate of energy deposition, whereas muons at T2K energies are minimum ionizing and 
are more longitudinally penetrating. 
The fourth variable is based on a region defined by only the two planes upstream and 
downstream of the event vertex and is useful for discriminating against showers with additional particles near the event vertex 
and proton-induced activity.
Since the total neutrino flux is constant and the neutral current (NC) cross section is the same for each neutrino flavor,
the NC event rate within INGRID is expected to be constant with sidereal time.
Accordingly, no additional cuts to remove NC events are used.
Figure~\ref{Likelihood} shows the $\mu_{CL}$ likelihood distribution for reconstructed data and Monte Carlo (MC) $\nu_{\mu}$ CC, $\nu_{e}$ CC and NC interactions.
%%%%%%%%%%%%%%%%%%%%%%%%%%%%%%%%%%%%%%%%5
%%
%   Muon Confidence Level distribution
%%
%%%
%
%%%%%%%%%%%%%%%%%%%%%%%%%%%%%%%%%%%%%%%%%%%%%%55
A cut on $\mu_{CL} \geq 0.54$ has been selected to ensure that the $\nu_{e}$ contamination of the final sample is smaller than  the statistical error on the $\nu_{\mu}$ component while maximizing the $\num$ statistics. After applying all analysis cuts the $\nu_{\mu}$ CC selection efficiency is  $\epsilon_{\mu} =44.0\%$. The corresponding $\nu_{e}$ efficiency, $\epsilon_{e}$, has been reduced to $13.3\%$. There are $6.75 \times 10^{6}$ events remaining in the final sample, which provides an average statistical error of $0.22\%$ in each of the 32 analysis bins (defined below). If an oscillation effect equivalent to three times the statistical error on the $\nu_{\mu}$ component appears as $\nu_{e}$ in the final sample the resulting contamination will be $0.2\%$. Assuming no oscillation due to LV effect, the final sample has $3.4\%$ NC events.

\subsection{Timing corrections and systematic uncertainties}
\label{sec:Corrections}
The operation of the T2K beam is not constant in time and varies with the hour of the  
day and season of the year. The effect of time-dependent changes in the neutrino event rate must be corrected since they can mimic an LV-oscillation signal or reduce the analysis sensitivity.
Such effects can be separated into two distinct classes depending on whether they alter the neutrino beam itself 
or the INGRID detector.
The first class consists of three time-dependent corrections considered for the neutrino beam:

\begin{itemize}

%neutrino beam center for each LSP bin within a given T2K run period
\item Beam center variations during each run:
 Since the neutrino interaction rate itself is insufficient to estimate these variations, muons 
 collected spill-by-spill with a muon detector just downstream of the decay 
 volume~\cite{PTEP2015} are used to estimate the beam center position.
 For each of the four run periods considered in this exposure, 
 the beam center position as a function of LSP is estimated 
 after correcting for tidal effects at the detector.
 These data are then used to extrapolate the position of the 
 neutrino beam center, which is aligned with the muon direction, at INGRID.
 LSP-dependent corrections to observed event rate at INGRID 
 due to shifts in the neutrino beam center are estimated 
 using MC.
  
\item Beam center variation between runs: 
Changes in the average beam center position between run periods are evaluated using the INGRID 
neutrino data and a correction is estimated and applied as in the above.

\item Beam intensity variation between runs and non-uniform POT exposure as a function of LSP:
  A correction is applied to bring the 
  event rate per POT in each LSP bin in line with the average 
  for the entire run. 
  The correction is applied for each event based on its run 
  and sidereal phase.
  A further correction is applied to make the average 
  event rate per POT of each run consistent with that 
  of a reference run chosen to be near the end of the 
  data taking period.

\end{itemize}

The second class of effects consists of three additional corrections to account for changes in the response of INGRID:

\begin{itemize}
\item Event pile-up variations:
  Typically only single interactions in an INGRID module are reconstructed and other interactions 
  in the same data acquisition timing window (one for each neutrino bunch) are lost (pile-up events).  
  However, changes in the beam intensity affect the probability of 
  multiple interactions within an INGRID reconstruction timing window.
  Accordingly, events at INGRID are corrected as a linear function of LSP
  to account for the variation in pile-up events with variations in the beam intensity.
  The number of lost pile-up events varies between 3\% and 7\% 
  across the INGRID modules.
  
\item Dark noise variations:
  Variations in the temperature and humidity 
  affect the MPPC dark rate, which in turn weakly affects the neutrino detection efficiency. 
  The maximal variations of the dark rate with the sidereal time is $2\%$. 
  A correction to account for this efficiency variation has been applied linearly 
  with the dark rate.

\item Variations in the photosensor gain:
  The MPPC gain is influenced by environmental changes, and the scintillator gain might decrease over time.
  Gain changes impact both the reconstruction and the analysis selection and are corrected using a sample of 
  beam-induced muon interactions in the rock upstream of INGRID.
  The effect of variations in the charge at the minimum ionization peak of these 
  muons is simulated in MC and used to correct the neutrino event rate. 
  The size of the correction varies with LSP and does not exceed 1\%.
\end{itemize}

\begin{figure}[hbt]
  \centering
  \includegraphics[width=.48\textwidth]{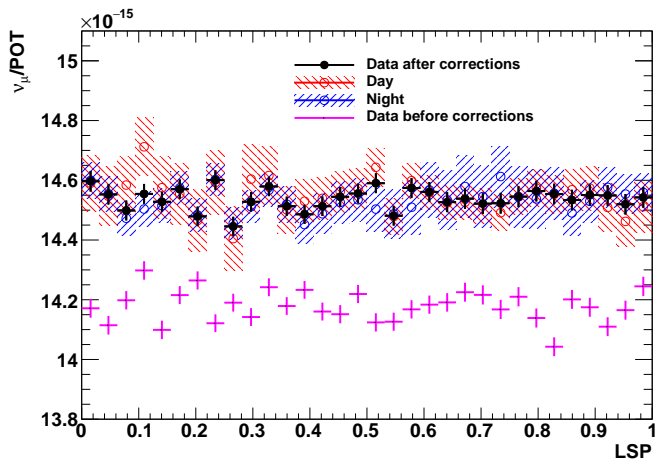}
  \caption{
           Distribution of reconstructed $\mu$-like events per POT as a function of LSP.
           Data before (magenta) and after (black) corrections are shown together 
           with the corrected sample additionally split into day (red) and night (blue) 
           subsamples.
           }
\label{BeforeAfterCorrections} 
\end{figure}

The validity of the above corrections has been tested by separating the analysis data set into day and night subsamples.
Though time-dependent differences are expected in the split samples due to, for instance, cooler 
temperatures at night or beamline maintenance during the day, the data should be consistent with one 
another when viewed in the LSP coordinate if the above corrections have been applied consistently.
Figure~\ref{BeforeAfterCorrections} shows the day and night distributions as a function of 
LSP. The agreement between the day and night distribution is evaluated with a Pearson’s chi-squared test and a corresponding $\chi^{2}/NDF = 28.3/32$ has been found.
Data before and after all corrections also appear in the figure.
Systematic errors for each of the corrections have been evaluated and are listed in Table~\ref{Tab:LV:Syssum}. The total systematic error is $0.08\%$, which is small when compared to the statistical error of the final sample, $0.22\%$.

%%%%
%%
%%  Corrected event rate per POT
%%
%%%%
%%
%%
%%

%%%%
%%
%% Systematic Errors for LSP corrections
%%
%%%
\begin{table}[ht]
\begin{center}
    \begin{tabular}{l|c}
      \hline
      \hline
      Source & Systematic uncertainty (\%) \\
      \hline
      \hline
      Pile-up                    & 0.01 \\
      MPPC dark noise      & 0.01 \\
      MPPC gain variation &  0.06   \\
      Beam position          & 0.03 \\
      Beam intensity & 0.05 \\
      \hline
      Total systematic & 0.08\\
      \hline
      \hline
    \end{tabular}
    \caption{\label{Tab:LV:Syssum}
      Summary of the $1\sigma$ systematic uncertainties induced from correcting for time dependent variations 
      in the neutrino event rate. The beam position variation between and within run periods have been combined into a single entry in the table.} 
\end{center}
\end{table}
%%
%%%

\section{Analysis methodology and results}

\begin{figure}[hbt]
  \centering
  \includegraphics[width=.48\textwidth]{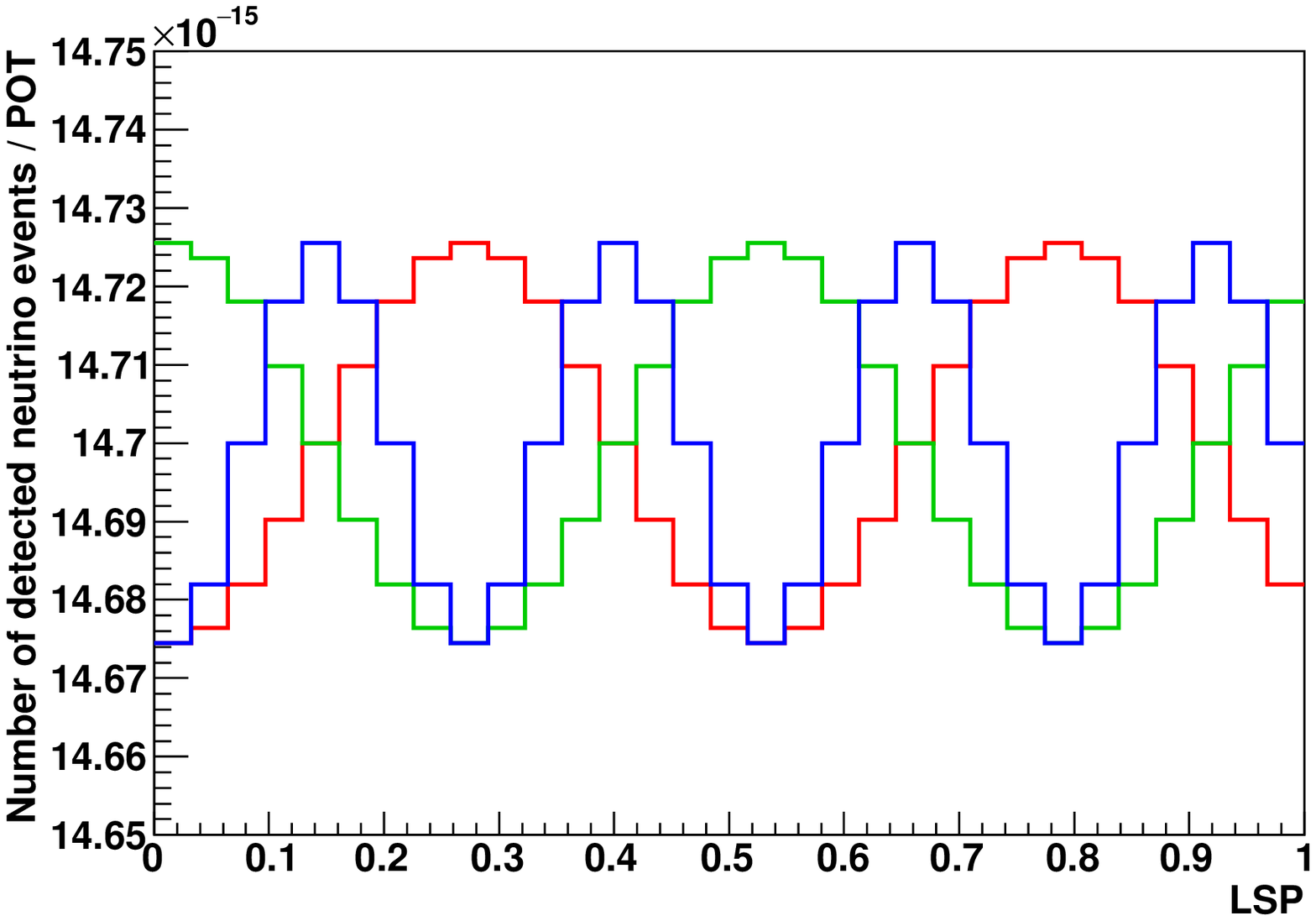}
  \caption{
           Distribution of the $\nu_{\mu}$ event rate 
           as a function of LSP for three different assumed signal configurations:
           $(\Ce,\Ace,\Ase,\Bce,\Bse)=(0,5\times10^{-20},0,0,0)~$GeV (red), $(0,0,5\times10^{-20},0,0)~$GeV (green), $(0,0,0,5\times10^{-20},0)~$GeV (blue). The coefficients corresponding to $\num \rightarrow \nut$ oscillation ($\Ctau,\Actau,\Astau,\Bctau,\Bstau$) have been set to 0.} 
\label{Signal} 
\end{figure}

The analysis of the final data sample is performed in two stages.
First, the compatibility of the data with a null signal is studied using a fast Fourier transform (FFT) method (Section~\ref{sec:FFT}).
This method explicitly searches for a sidereal modulation and ultimately provides an estimate of the power 
of each Fourier mode from a potential signal.
Then, constraints on the parameters appearing in Equation~(\ref{eq:shortprob}) are extracted 
using a likelihood method (Section~\ref{sec:Likelihood}) that includes their correlations.
%%Both methods have been tested on toy MC experiments with and without LV oscillations.
Figure~\ref{Signal} shows examples of the expected LSP distribution for MC generated under three signal assumptions.

%%%%
%%
%   Toy MC Image
%%
%%%%
%%%%%
%%%%%

\subsection{The Fast Fourier transform result}
\label{sec:FFT}

Expanding Equation~(\ref{eq:shortprob}) indicates that LV oscillations are described by four harmonic sidereal frequencies 
$f_{i} = i \cdot \somega$, $i \in [1,4]$ and a constant term.
The FFT~\cite{press2007numerical,duhamel1990fast} method is most efficient for $N = 2^{L}$ bins 
and the sensitivity of the current analysis is found to be optimal when $L= 5 $.
Data are therefore divided into 32 evenly spaced LSP bins for input into the FFT 
and the magnitudes of the four Fourier modes, $|F_{i}|$, are then estimated.
Note that the constant term is not considered in this study due to large uncertainties in the beam flux normalization.
A $3 \sigma$ detection threshold has been determined as the power in a Fourier mode for which 
0.3\% of MC experiments generated without LV effects shows higher power.    
For each mode this threshold corresponds to $|F_{i}| > 0.026$.
The results of the fit to the data are shown in Table~\ref{Tab:FFTresults} together with a p-value 
estimating the likelihood that the observed power was produced by a statistical fluctuation of the null (no LV) hypothesis.
All $|F_{i}|$ are below the $3\sigma$ detection threshold and indicate no evidence for a LV signal.

%%%%%%%
%%%
%%    Magnitudes from Fourier Analysis  
%%%
%%%%%%%
\begin{table}[ht]
\caption{ Observed power in each Fourier mode from a fit to the data using the FFT method.
          A positive observation at $3\sigma$ would correspond to an observed power greater than 0.026 
          in any $\somega$.}
\begin{center}
    \begin{tabular}{l|c|c}
      \hline
      \hline
      Fourier Mode & Magnitude & p-value \\
      \hline
      \hline
      $|F_{1}|$  & 0.011 & 0.35 \\
      $|F_{2}|$  & 0.009 & 0.48 \\
      $|F_{3}|$  & 0.006 & 0.69 \\
      $|F_{4}|$  & 0.009 & 0.51 \\
      \hline
      \hline
    \end{tabular}
\end{center}
\label{Tab:FFTresults}
\end{table}
%%%%%%%%%%%%%%%%%%

\begin{figure}[hbt]
  \centering
  \includegraphics[width=.45\textwidth]{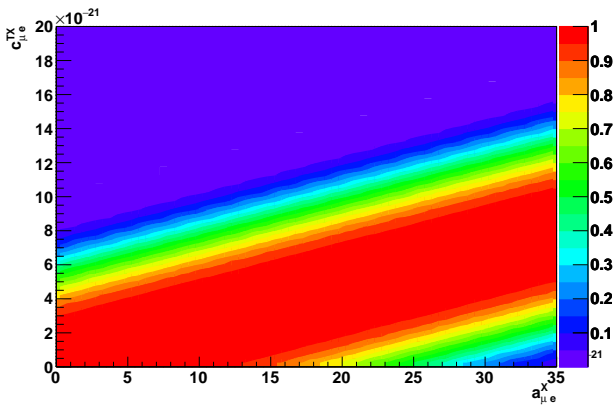}
  \caption{
           Probability for the observed Fourier power in a null observation to exceed the expected power from 
           a LV signal as a function of the $(a_L)^{X}_{\mu e}$ and $(c_L)^{TX}_{\mu e}$ coefficients.} 
           \label{Fig:LV:Correlations} 
\end{figure}

Constraints on the SME coefficients can be extracted with the FFT method~\cite{Adamson:2008aa,diaz2009perturbative} under the assumption that the parameters above are uncorrelated.
However, since the data sets are reduced to the four amplitudes and the relatively large number of parameters in the 
oscillation function, correlations are expected.
Figure~\ref{Fig:LV:Correlations} shows the probability for data without LV 
to yield more power in the Fourier modes than the average expected for a LV signal as a function 
of the SME coefficients $(a_{L})^{X}_{\mu e}$ and $(c_{L})^{TX}_{\mu e}$.
The parameters exhibit a high degree of anti-correlation, indicating that in the event of a null observation 
as above, using the FFT method without considering these correlations may lead to an underestimation of the 
parameter limits.
As the parameters in Equation~(\ref{eq:shortprob}) are functions of these coefficients, they might be also 
expected to exhibit correlations. 
Accordingly, a likelihood method has been developed to fully incorporate these correlations when
making parameter estimations.

%%%%%%%%%%%%%%%%%
%%%
%%  Parameter Correlations
%%%
%%%%%%
%%%%%%%%%%%%%%%%%

\subsection{Likelihood analysis}
\label{sec:Likelihood}

Due to the large number of SME parameters~\cite{diaz2009perturbative} relative to the number of observables, 
this analysis does not estimate the $(a_L)^{X}_{ab}$ and $(c_L)^{TX}_{ab}$ parameters but the $\Cmb$, $\Acmb$, $\Asmb$, $\Bcmb$, $\Bsmb$ ($b=e,\tau$) parameters from 
Equation~(\ref{eq:shortprob}) using a likelihood method that fully incorporates 
their correlations and the experimental uncertainties.
However, since the impact of systematic errors is negligible (c.f. Table~\ref{Tab:LV:Syssum}), only the statistical uncertainty in each LSP 
bin is considered here.
Further, each parameter is assumed to be real valued.
Sensitivity studies without this assumption showed no  
significant constraint on the complex phases of these parameters with the present data.
Under these conditions, a simultaneous fit for ten real parameters using the data and binning from the previous section has been 
performed. Since the parameters are highly correlated, the contours and limits are not estimated assuming a profiling method, but instead using a likelihood marginalization which genuinely preserve their correlations~\cite{PDGstat}. This analysis assumes flat priors for all the parameters since no LV has been discovered so far. The results of the fit are shown in the Table~\ref{Tab:LV:FitResults}. 

%%%%%%%%%%%%%%%%%%
%%
%   Best fit and Systematic errors for 5 coeff. analysi
%%
%%
  \begin{table}[htb]
    %\begin{center}
    %\footnotesize
    \caption{\label{Tab:LV:FitResults}
      Best fit (BF) values with $68\%$, and  $95\%$ upper
      limit values on the LV model parameters using the likelihood method (in units of $10^{-20}$ GeV). In the last row, the expected sensitivity is shown.}
    \begin{tabular}{l|c|c|c|c|c}
      \hline
      \hline
      & $\Ce$ & $\Ace$ & $\Ase$ & $\Bce$ & $\Bse$ \\
      \hline
      Best fits & -0.3 & 0.3 & 0.4 & -1.2 & 2.0 \\
      $68\%$ C.L Limits & 1.3 & 1.5 & 2.0 & 1.3 & 1.6 \\
      95\% C.L Limits & 3.0 & 3.2 & 3.8 & 2.6 & 3.1 \\
      95\% C.L Sensitivity & 2.5 & 2.7 & 4.3 & 3.5 & 3.5 \\
      \hline
      & $\Ctau$ & $\Actau$ & $\Astau$ & $\Bctau$ & $\Bstau$ \\
      \hline
      Best fits & -0.8 & -0.4 & -3.2 & -0.4 & 1.1 \\
      $68\%$ C.L Limits & 1.3 & 1.5 & 2.0 & 1.3 & 1.6 \\
      95\% C.L Limits & 3.0 & 3.2 & 3.8 & 2.6 & 3.1 \\
      95\% C.L Sensitivity & 2.5 & 2.7 & 4.3 & 3.5 & 3.5\\
      \hline
      \hline
    \end{tabular}
    %\end{center}
  \end{table}
%%%%%%%%%%%%%%%%%%

\begin{figure}[htb]
  \centering
  %\subfigure{\includegraphics[width=.22\textwidth]{Contour_0_1.eps}}
  \includegraphics[width=.5\textwidth]{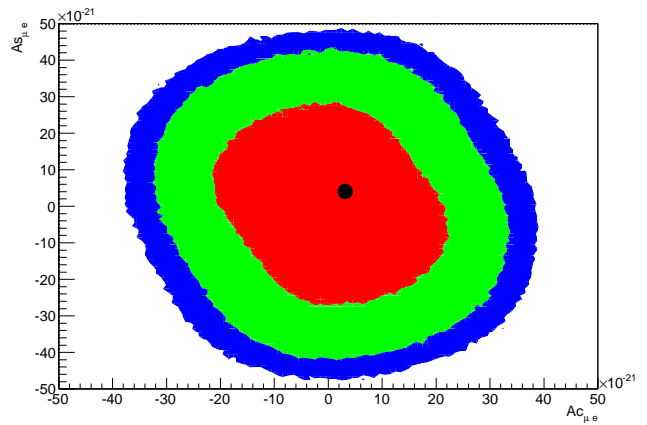}
  \caption{Ten-coefficient fit result in the $\Ace,\Ase$ coefficient plane. The other parameters are marginalized over. The best fit point is marked in black, with $68\%$, $90\%$ and $95\%$ credible intervals shown in red, green and blue, respectively.} 
  %\caption{Ten-coefficient fit result in the $\C,\Ace$ (left) and $\Ace,\Ase$ (right) coefficient planes. The other parameters are marginalized over. The best fit point is marked in black, with $1\sigma$ and $2\sigma$ contours shown in red and blue, respectively.} 
\label{Fig:LV:AcAs}
\end{figure}

As expected from the FFT method, no indications of LV oscillations are found and $2 \sigma$ upper limits are set for each parameter. Those limits are compared with the sensitivity obtained by determining the parameter absolute values for which $5\%$ of some MC experiments generated without LV effects shows higher absolute values. The contour limits are constructed following a constant $\Delta \chi^{2}$ method and are shown in Figure~\ref{Fig:LV:AcAs} for the $\Ace$ and $\Ase$ parameters that show important anti-correlations.
While correlated-parameter analyses have been performed elsewhere~\cite{Katori:2012pe}, this is the first search 
to do so using all ten parameters simultaneously. The five harmonics in Equation~(\ref{eq:shortprob}) heavily correlate the ten parameters as shown in Figure~\ref{Fig:LV:AcAs}. Neglecting the correlations between the parameters will lead an underestimation of the parameter limits. Since these correlations vary with the direction and position of each experiment, any comparison or combination of the limits found by different experiments requires to preserve these correlations.
%A unique feature of this analysis is its ability to disentangle sign degeneracies among parameters as depicted in Figure~\ref{Fig:LV:AcAs}. Since preserving the correlations between the $\Ce$, $\Ctau$,$\Ace$, $\Actau$,$\Ase$, $\Astau$,$\Bce$, $\Bctau$ and $\Bse$, $\Bstau$ their sign can be constrained through their cross-products coming from the expansion of Equation~\ref{eq:shortprob}. In addition this result gives the first constraints on $\Bce$ and $\Bse$ within the context of a correlated ten parameter fit.

%%%%%%%%%%%%%%%%%%
%%
%   Best fit and Systematic errors for 5 coeff. analysi
%%
%%
%%%%%%%%%%%%%%%%%%

\section{Conclusions}
The T2K experiment has performed a search for Lorentz and CPT invariance violations using the INGRID on-axis near detector.
Two complementary analysis methods have found no evidence of such symmetry violations for the energy, 
neutrino baseline, and data set used.
Not only are the data consistent with an LSP-independent event rate based on a FFT analysis,
but a likelihood analysis incorporating parameter correlations has corroborated this finding and yielded constraints on ten SME parameters.\\
\\
\section*{Acknowledgements}
We thank the J-PARC staff for superb accelerator performance. We thank the 
CERN NA61 Collaboration for providing valuable particle production data.
We acknowledge the support of MEXT, Japan; 
NSERC (Grant No. SAPPJ-2014-00031), NRC and CFI, Canada;
CEA and CNRS/IN2P3, France;
DFG, Germany; 
INFN, Italy;
National Science Centre (NCN) and Ministry of Science and Higher Education, Poland;
RSF, RFBR, and MES, Russia; 
MINECO and ERDF funds, Spain;
SNSF and SERI, Switzerland;
STFC, UK; and 
DOE, USA.
We also thank CERN for the UA1/NOMAD magnet, 
DESY for the HERA-B magnet mover system, 
NII for SINET4, 
the WestGrid and SciNet consortia in Compute Canada, 
and GridPP in the United Kingdom.
In addition, participation of individual researchers and institutions has been further 
supported by funds from ERC (FP7), H2020 Grant No. RISE-GA644294-JENNIFER, EU; 
JSPS, Japan; 
Royal Society, UK; 
and the DOE Early Career program, USA.
%\bibliographystyle{plainnat}
%\bibliography{Bibnew.bib}


\end{document}